\documentclass[preprint,number,sort&compress,longtitle]{elsarticle}   
\pdfoutput=1 

\journal{Nuclear Instruments and Methods in Physics Research A}

\usepackage[switch]{lineno}

\usepackage[english]{babel}
\usepackage{times}
\usepackage{graphicx}
\usepackage{fontenc}
\usepackage{mathptmx}
\usepackage{subfigure}
\usepackage{amsmath,amssymb}
\usepackage{upgreek}
\usepackage{xspace}
\usepackage{color}
\definecolor{darkred}{rgb}{0.5,0,0}
\definecolor{darkblue}{rgb}{0,0,0.5}
\definecolor{firebrick}{rgb}{0.75,0.125,0.125}
\definecolor{darkgreen}{rgb}{0,0.5,0}
\usepackage[pdftex,colorlinks=true]{hyperref}
\usepackage{units}
\usepackage{booktabs}
\usepackage{epstopdf}
\usepackage{amsmath}

\graphicspath{{plots/}}
\makeatletter
\def\elsartstyle{%
   \def\normalsize{\@setfontsize\normalsize\@xiipt{14.5}}
   \def\small{\@setfontsize\small\@xipt{13.6}}
    \let\footnotesize=\small
    \def\large{\@setfontsize\large\@xivpt{18}}
    \def\Large{\@setfontsize\Large\@xviipt{22}}
    \skip\@mpfootins = 18\p@ \@plus 2\p@
    \normalsize
}
\@ifundefined{square}{}{}
\makeatother

\makeatletter
\def\blfootnote{\xdef\@thefnmark{}\@footnotetext}
\makeatother

\begin{document}

\begin{frontmatter}

\title{GIGAS: a  set of microwave sensor  arrays to detect  molecular bremsstrahlung  radiation from extensive air shower}
\author[1]{R.~Ga\"{i}or\corref{cor1}}
\ead{romain.gaior@lpnhe.in2p3.fr}
\author[1,2]{I.~Al Samarai\fnref{imen}}
\author[3]{C.~B\'{e}rat}
\author[1]{M.~Blanco Otano}
\author[1]{J.~David}
\author[2]{O.~Deligny}
\author[1]{H.~Lebbolo}
\author[3]{S.~Lecoz\fnref{sandra}}
\author[1]{A.~Letessier-Selvon}
\author[2]{I.~Lhenry-Yvon}
\author[1]{I.~C.~Mari\c{s}\fnref{ioana}}
\author[3]{F.~Montanet}
\author[1]{P.~Repain}
\author[2]{F.~Salamida\fnref{francesco}}
\author[1]{M.~Settimo\fnref{mari}}
\author[3]{P.~Stassi}
\author[3]{A.~Stutz}

\cortext[cor1]{Corresponding author}
\fntext[imen]{Now at DPNC
                                Universit\'{e}       de      Gen\`{e}ve,
                                Switzerland}
\fntext[sandra]{Now at National
   Astronomical  Observatory,  Chinese  Academy  of  Sciences,  Beijing
   100012, China}              
\fntext[ioana]{Now at Universit\'{e}  Libre  de
                                 Bruxelles (ULB), Brussels, Belgium}     
\fntext[francesco]{ Universit\`{a} dell'Aquila,
   Dipartimento di Scienze Fisiche e Chimiche, L'Aquila, Italy}
\fntext[mari]{Now at SUBATECH, Universit\'{e}  de Nantes, Ecole des
   Mines de Nantes , CNRS-IN2P3, Nantes, France}


\address[1]{Laboratoire de Physique Nucl\'{e}aire et de Hautes Energies (LPNHE), Universit\'{e}s
Paris 6 et Paris 7, CNRS-IN2P3, Paris,
France}
\address[2]{Institut de Physique Nucl\'{e}aire d'Orsay (IPNO), Universit\'{e} Paris 11, CNRS-IN2P3,
France}
\address[3]{Laboratoire de Physique Subatomique et de Cosmologie (LPSC), Universit\'{e}
Grenoble-Alpes, CNRS-IN2P3,
France}

\begin{abstract}
  \noindent

We present  the GIGAS (Gigahertz  Identification of Giant  Air Shower)
microwave  radio sensor arrays  of the  EASIER project  (Extensive Air
Shower Identification with Electron Radiometers), deployed at the site
of the  Pierre Auger cosmic ray  observatory.  The aim  of these novel
arrays  is to  probe  the intensity  of  the molecular  bremsstrahlung
radiation expected  from the development of the  extensive air showers
produced by  the interaction of ultra  high energy cosmic  rays in the
atmosphere.  In  the designed setup,  the sensors are  embedded within
the surface detector array of the Pierre Auger observatory allowing us
to  use the  particle signals  at ground  level to  trigger  the radio
system.  A series of seven, then  61 sensors have been deployed in the
C-band, followed by a new series  of 14 higher sensitivity ones in the
C-band and the L-band.  The design, the operation, the calibration and
the sensitivity to extensive air showers of these arrays are described
in this paper.
\end{abstract}
\end{frontmatter}
\setcounter{footnote}{0}
\section{Introduction}
The very low flux of Ultra High Energy Cosmic Rays (UHECRs), less than
one  particle  per  year  per square  kilometers  above  10$^{19}$~eV,
requires a very large detection surface at ground level to measure the
primary particle  properties via the Extensive Air  Showers (EAS) they
produce in  the atmosphere.   Such an indirect  measurement is  a very
difficult  technical challenge and  the search  for efficient  but low
cost techniques is an ongoing process.  \\

As of  today, the largest  cosmic ray observatories, the  Pierre Auger
Observatory~\cite{augerdetector}    and   Telescope   Array~\cite{TA},
combine two techniques to measure  EAS: an array of particle detectors
at  ground  supplemented  by  a  network  of  fluorescence  telescopes
overlooking  the  atmosphere above  the  ground  array.  The  particle
detectors,  scintillators  or water  Cherenkov  detectors, sample  the
lateral  profile of  the cascades  that  reach the  ground, while  the
telescopes  measure  the  longitudinal  profile  of  the  cascades  by
detecting the fluorescence light emitted isotropically by the nitrogen
atoms  previously  excited  by  the  passage  of  the  electromagnetic
component of  the EAS.  While  extremely powerful, the  combination of
these two  techniques suffers from  the limited duty cycle,  less than
15\%, of the  fluorescence technique, which can only  be active during
clear  moonless nights.   With such  a  limited duty  cycle, the  mass
composition information,  best determined by the depth  of the maximum
of the  longitudinal profiles,  is essentially unavailable  for UHECRs
with energies above  $\simeq4\times10^{19}$~eV.  As such, the question
of the origin and nature of UHECR in this energy range, where a strong
flux  suppression has been  measured in  the energy  spectrum, remains
unsolved.  Indeed,  the interpretation of this cut-off  is still being
debated~\cite{Fukushima:2013yea}. \\

One of  the main  motivation for  using the radio  emission of  EAS to
measure UHECR properties relies on its capabilities to provide similar
information as  the fluorescence technique but without  the duty cycle
limitation.     Initially   proposed    and    implemented   in    the
1960's~\cite{jelley65},  the   radio  detection   of  EAS  is   now  a
well-established technique  and has been  mostly exploited in  the VHF
band~\cite{codalema,  lopes, aera2}  and~\cite{huegeradioreview}  for a
recent  review.  In this  frequency range,  the observed  radiation is
mainly produced by the acceleration  of the electrons of the shower in
the geomagnetic field, and, to  a smaller extent, by the moving charge
excess (also know  as Askaryan radiation)~\cite{augerradio}.  However,
both of  these radiations  are beamed forward  in the  Cherenkov cone,
which  is around  $\simeq 1^\circ$  in  air, and  centered around  the
shower  axis.  The  resulting imprint  of the  radio signal  at ground
level  is  generally observable  only  up  to  a few  hundred  meters,
limiting this technique  to densely-instrumented arrays. Although such
densely-instrumented arrays can be deployed over surfaces that provide
good sensitivity to  study cosmic rays with primary  energies of about
\unit[$10^{18}]{eV}$ or  less, the cost  is then a limiting  factor to
envisage the deployment of antennas  over the surfaces needed to probe
the  flux  at  the  highest  energies.  In  addition,  the  collimated
emission  of   the  radiation  limits  the  ability   to  measure  the
longitudinal development of the showers as one only detects the ground
projection of this profile around the shower axis.  \\

In 2008,  an accelerator experiment,  SLAC T471, detected a  signal in
the  microwave frequency  range (1.5-6  GHz)  upon the  passage of  an
electromagnetic  shower in  an  anechoic chamber~\cite{Gorham}.   This
signal  was interpreted as  Molecular Bremsstrahlung  Radiation (MBR),
and its  intensity, extrapolated to UHECR energies,  was then expected
to be  detectable with rather simple radio-detector  systems.  The MBR
is produced  by the  acceleration of the  ionization electrons  in the
electric  field of  the  atmosphere molecules.   The  radiation is  in
principle isotropic  and its intensity directly related  to the energy
deposited by the EAS particles  in the atmosphere.  MBR profiles would
therefore  be very similar  to fluorescence  ones, with  the advantage
that  MBR detection  in the  microwave band,  where the  atmosphere is
essentially transparent, can  be done with a 100\%  duty cycle. Such a
promising signal, together with the fact that sensors in that band are
very cheap due to their commercial use for satellite TV reception, led
to  the development  of additional  accelerator experiments  to better
characterize the signal~\cite{amy},~\cite{maybe}. In addition, several
in  situ experiments  aiming  at  the direct  observation  of the  MBR
emission  from EAS like  MIDAS~\cite{midas}, CROME~\cite{cromeresult},
AMBER~\cite{amber} and EASIER were also set up. \\

The  combined measurements  of  the EAS  longitudinal  profile in  the
atmosphere  and   of  the  particle  contents  at   ground  allow  the
reconstruction  of  the  shower  key  parameters  and  thus  a  better
understanding of the  mass of UHECRs.  For a  large enough MBR signal,
the EASIER  setup, which is presented  in this paper,  was designed to
allow an access  to several composition indicators such  as the shower
depth of  maximum and the muonic-to-electromagnetic ratio  on an event
by event basis.  This information would help to understand whether the
suppression  observed in  the energy  spectrum  is the  result of  the
extinction  of  the  sources  (i.e. the  acceleration  mechanisms  has
reached its maximum potential) or  the result of a propagation effect,
due to the interaction of  UHECRs with the cosmic microwave background
of    radiation   (the    Greisen,   Zatsepin    and    Kuzmin   (GZK)
cut-off~\cite{gzk1,gzk2}). \\

In this  paper, we present the  developments of the  EASIER project, a
concept of radio detectors integrated  to the Surface Detector (SD) of
the Pierre  Auger Observatory. An  EASIER detector is a  radio antenna
combined with an envelope detector integrated to an SD station. EASIER
thus take advantage of the power supply and data acquisition but takes
most  of  its benefits  from  the  station  trigger.  Radio  Frequency
Interference (RFI) and especially the anthropogenic noises are by this
mean filtered out.   Thus, this setup has the  capability to probe the
radio signal from UHECRs at  large distances from the shower axis.  In
section~\ref{sec:setup},  the  general  concept  of  EASIER  is  first
presented  prior the description  of the  three different  versions of
microwave  sensor arrays  installed in  the GHz  frequencies: GIGAS61,
GIGADuck-C and GIGADuck-L.  The full calibration of these detectors is
then detailed in section~\ref{sec:calibration}.  Finally the method to
simulate  the  MBR  is  described  and  combined  to  the  calibration
information  to  produce estimations  of  the  systems sensitivity  in
section~\ref{sec:simulation}.

\section{The EASIER detection setup, GIGAS61 and GIGADuck detectors}
\label{sec:setup}
EASIER is  a novel radio-detector  concept composed of a  radio sensor
and of  an envelope  detection electronics embedded  in the SD  of the
Pierre  Auger  Observatory.  This  concept  was  implemented in  three
bandwidths: the VHF  band (30-80~MHz), the L band  (1-1.5~GHz) and the
C-band (3.4-4.2~GHz).  We  focus in this article on  the L- and C-band
only. The EASIER  experiment is one of the  three experiments deployed
at the Pierre  Auger Observatory to search for the  MBR emitted by the
ionization electrons left  in the atmosphere after the  passage of the
shower. In  contrast to the two other  ones, namely AMBER~\cite{amber}
and  MIDAS~\cite{midas},  which  instrument  an  array  of  feed  horn
antennas  illuminated  by  a  parabolic  dish, EASIER  relies  on  the
observation  of the shower  from the  ground level  with a  wide angle
antenna  pointing directly  to the  sky.  In  2011, a  first set  of 7
antennas was  deployed, followed by  54 additional in 2012  making the
\mbox{GIGAS61}  array.   The  analysis  of  \mbox{GIGAS61}  data  has  revealed  the
observation    of   radio    signals   emitted    by   EAS    in   the
C-band\cite{amber}.  However,  such detection  occurred  only for  air
showers at  distances less than around \unit[200]{m}  from the \mbox{GIGAS61}
antenna and could  be also explained by other  emission processes than
the   MBR.   Furthermore,   new  estimations   of  the   expected  MBR
intensity~\cite{imen2015,imen2016} led  to the development  of two new
versions of  EASIER, called \mbox{GIGADuck-C}  (installed in March  2015) and
GIGADuck-L (installed in December  2016), with an enhanced sensitivity
to search for signal from larger distances hence fainter.

\subsection{The electromagnetic background at the Pierre Auger Observatory}
\begin{figure}[!t]
  \center {
    \subfigure{\includegraphics[width=0.49\linewidth]{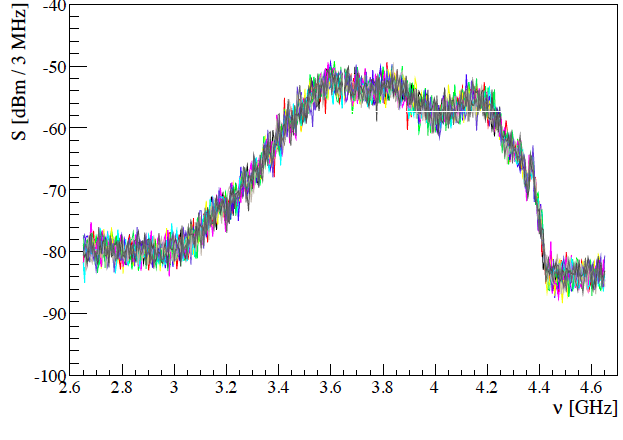}}
    \subfigure{\includegraphics[width=0.49\linewidth]{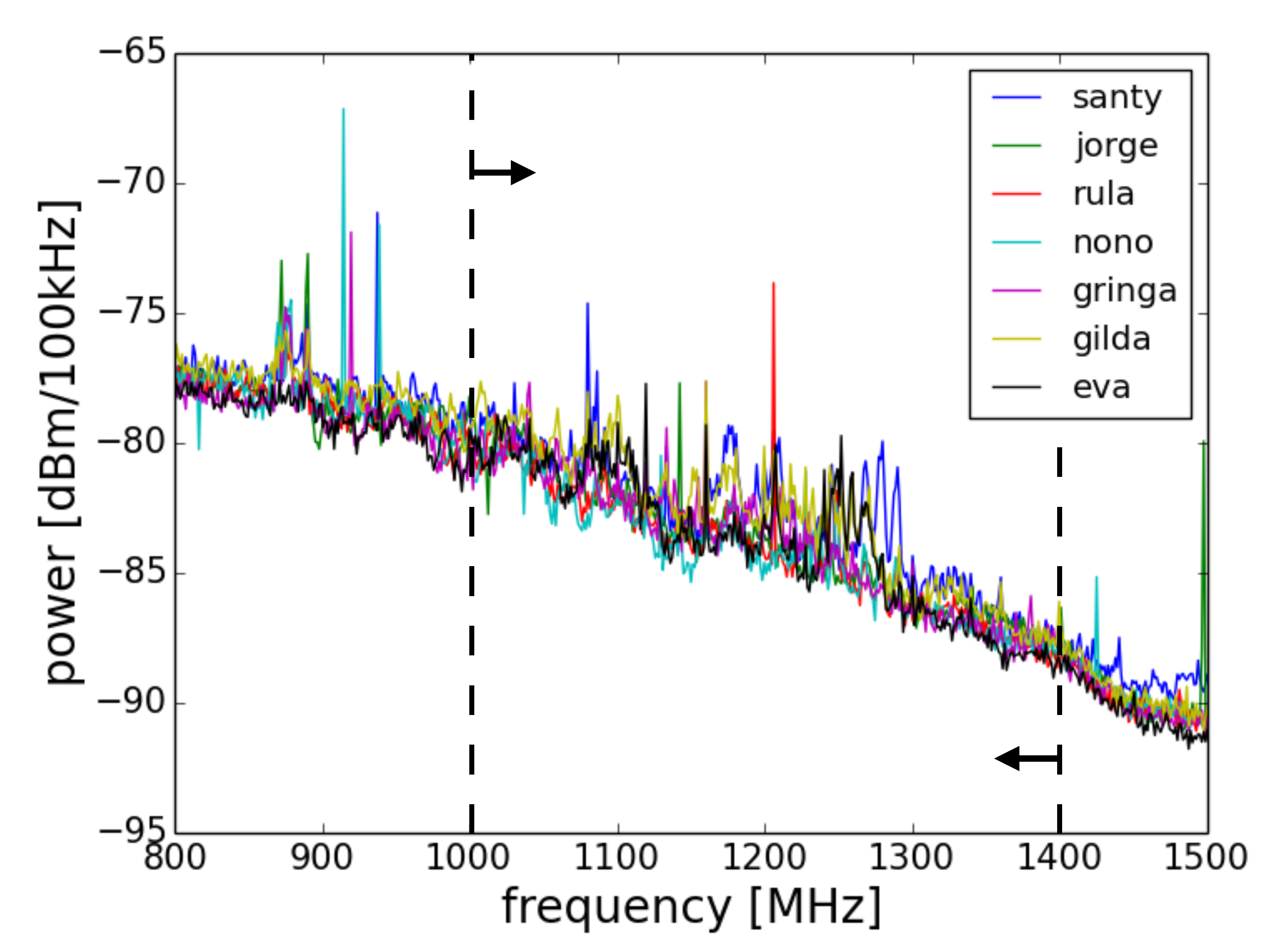}}
    \caption{\small{Frequency spectra in  Pampa Amarilla in the C-band
        recorded (left)  and in the  L-band (right). The  band between
        the  dashed line  is the  frequency  band once  the filter  is
        applied.}
      \label{fig:Pampa_spec}
    }
}
\end{figure}

Radio measurement are often hindered  by man-made noise.  Prior to the
installation, the electromagnetic background  was measured on the site
of the Pierre  Auger Observatory located in the  Pampa Amarilla in the
province of  Mendoza in Argentina.   Figure~\ref{fig:Pampa_spec} shows
the  power spectrum between  \unit[2.6 and  4.6]{GHz} measured  with a
C-band LNBf (Low Noise Block feed).  The gain of the amplifier used is
roughly \unit[60]{dB} between \unit[3.4 and 4.2]{GHz}. With a recorded
power in the sensitive  bandwidth of \mbox{\unit[-55]{dBm / 3MHz}} the
noise  floor is  thus about  \unit[-180]{dBm/Hz}.  No  strong  peak is
observed above this level in  the tens of recorded spectra making this
band   adequate   for   our   experiment.\\   In   the   L-band   (see
Figure~\ref{fig:Pampa_spec}  (right)),   peaks  can  be   noticed,  in
particular  around \unit[900]{MHz}  where a  strong  intermittent peak
could  be observed.   It originates  from the  Auger  SD communication
system  and  the  mobile  phone  band  and  can  be  reduced  with  an
appropriate  filtering.   Other  peaks  are also  present  inside  the
frequency  band of interest  between \unit[1  and 1.4]{GHz}  but their
amplitude remains acceptable.

\subsection{The experimental setup}  
\begin{figure}[!ht]
  \centering
  \hspace*{-3ex}  
\subfigure{\includegraphics[width=0.59\linewidth]{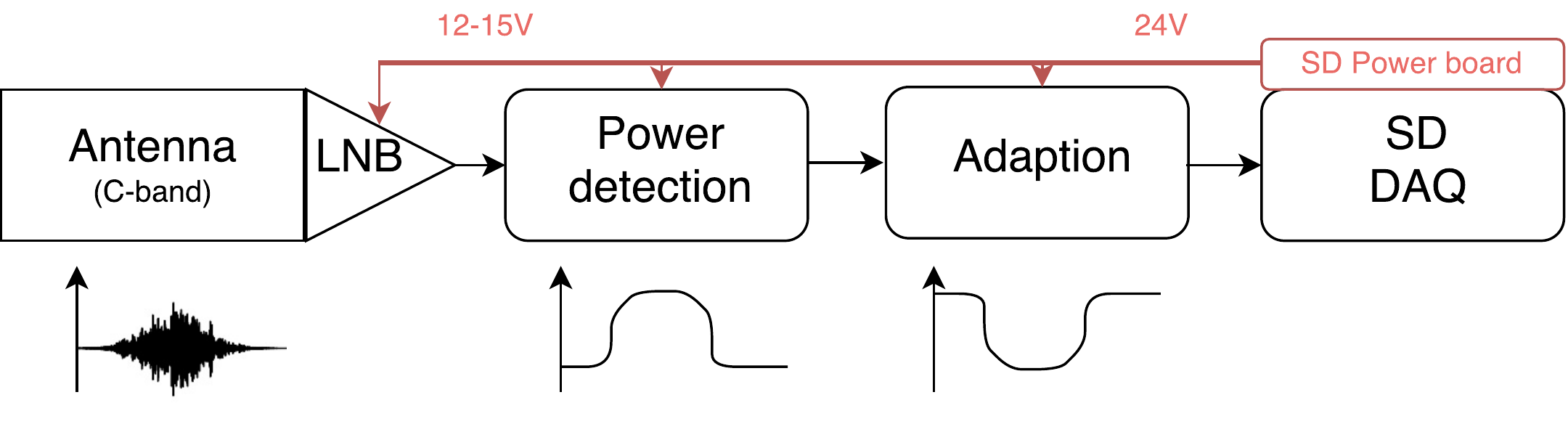}}  
  \caption{Block diagram of an EASIER detector unit.}
 \label{fig:diagram}  
\end{figure}

\begin{figure}[!ht]
  \centering
  \hspace*{-3ex}  
    \subfigure{\includegraphics[width=0.69\linewidth]{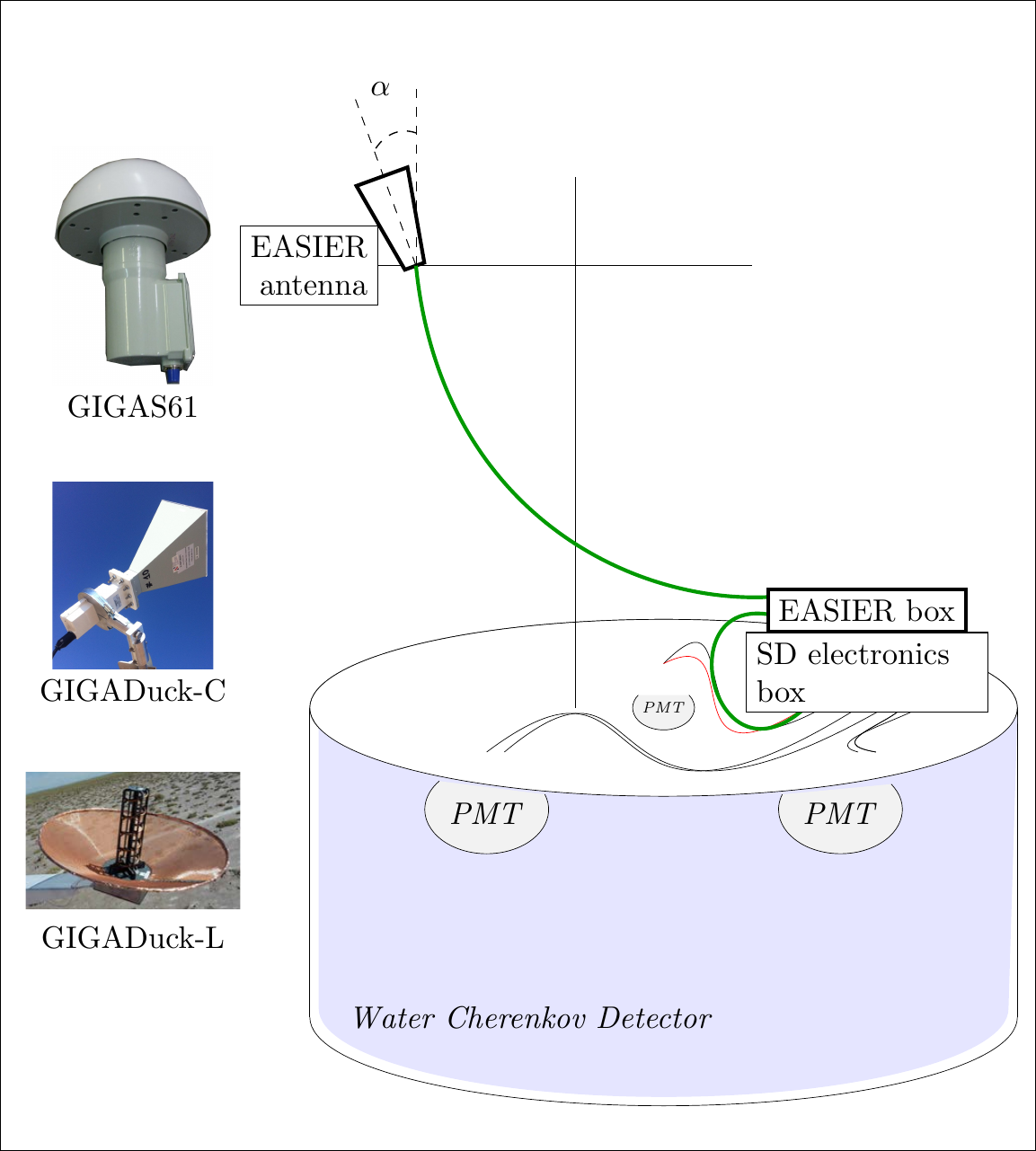}}  
  \caption{EASIER general  scheme. One of the three  antennas shown on
    the left  hand side is installed on  a pole that sits  on the WCD.
    The antenna  is vertical  in the case  of \mbox{GIGAS61}  and 6 out  of 7
    antennas are  tilted by  an angle $\alpha$  = 20$^{\circ}$  in the
    GIGADuck hexagon design (see text). The RF signal is amplified and
    transferred to the GIGAS box to be transformed in its envelope and
    acquired in the SD acquisition.}
 \label{fig:easierscheme}  
\end{figure}

The EASIER detector is embedded in a sub-array of the surface detector
(SD) of the Pierre Auger Observatory. The SD is composed of 1660 water
Cherenkov  detectors (WCD)  arranged in  a triangular  grid  of 1500~m
spacing. Each WCD is equipped with three Photo Multiplier Tubes (PMT),
a     local     acquisition     and    a     communication     system,
see~\cite{augerdetector}  for  a   detailed  description.   An  EASIER
detector unit is designed to be  integrated into a WCD. It is composed
radio sensor installed on top of the SD station and an electronics box
located  below  the  hatch  box  on  top  of  the  SD  electronics(see
Fig.~\ref{fig:easierscheme}). The three setups described in this paper
share  common elements presented  rightafter, their  specificities are
addressed in the following paragraphs.\\ The sensor is an antenna with
a main  lobe of  30 to 45$^\circ$  depending on the  considered setup.
Since the  expected radiation is unpolarized, there  is no requirement
on the polarization type of the antenna.  The sensor is followed by an
amplification and  a filtering stage.   The radio frequency  signal is
then  transformed into  a power  envelope by  a  logarithmic amplifier
(Analog Device  AD8318) which delivers  a voltage proportional  to the
logarithm of the RF input power.   This model was chosen for its large
frequency bandwidth  and its  fast time response  of a  few \unit[tens
  of]{ns}.  The output voltage is in  turn adapted to the front end of
the WCD electronics which is originally built to accept PMT's negative
voltage   between  0  and   -2V  (see   Fig.~\ref{fig:diagram}).   The
adaptation is performed through an amplification that sets the dynamic
range  to \unit[20]{dB}  and an  offset  used to  adjust the  baseline
level.  The EASIER analogic signal replaces one of the six channels of
the  WCD front  end electronics.   The final  part of  the acquisition
includes   an   antialiasing    filter   cutting   frequencies   above
\unit[20]{MHz}  and  the  FADC  (Flash Analog  to  Digital  Convertor)
digitizer.  The recorded waveform is \unit[19.2]{$\mu$s} long acquired
with a \unit[40]{MS/s} rate and has an amplitude sampled over 1024 ADC
units (refereed  as ADCu in  the following)~\cite{augerdetector}.  The
data  stream  is  then  sent   to  the  central  acquisition  and  the
reconstruction  of the  EAS event  is performed  independently  of the
radio signals.   As a consequence,  no separate trigger for  the radio
signal is needed and the EASIER data are simply part of the regular SD
data stream.  As an additional  benefit, the radio detector is powered
by  the station battery  and is  also integrated  into the  SD station
monitoring system.
\subsubsection*{GIGAS61}
\begin{figure}[!t]
  \centering
  \hspace*{-3ex}  
  \subfigure{\includegraphics[width=0.45\linewidth]{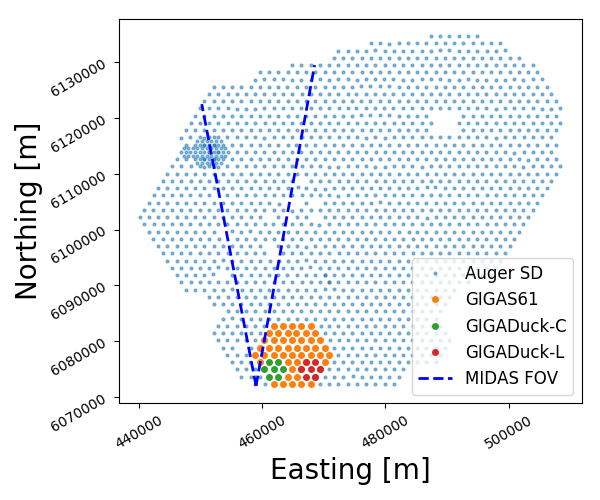}}
  \subfigure{\includegraphics[width=0.54\linewidth]{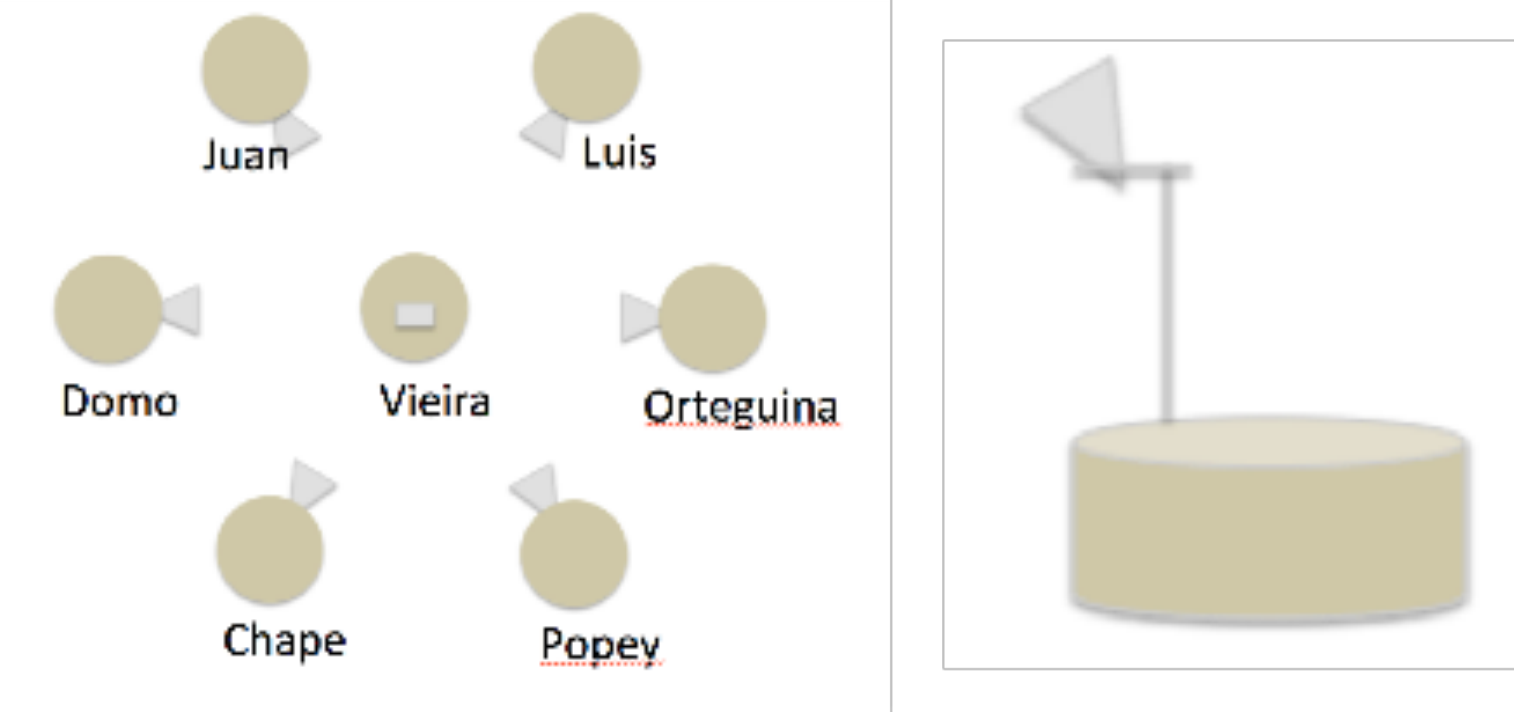}}
  \caption{Left: GIGAS61, and GIGADuck arrays layout within the Pierre
    Auger SD.  Middle and right: Top  view of GIGADuck  array and side
    view of one detector.}
  \label{fig:detector}  
\end{figure}

The    \mbox{GIGAS61}    antenna     is    a    commercial    horn    antenna
(Fig.~\ref{fig:easierscheme}) made of a cylindrical feed and a quarter
wave length monopole at its  bottom. The metallic ring around the feed
reduces the backlobe and widens the main lobe.  A hemispherical radome
is glued  to the ring to  protect the antenna from  rain.  The antenna
has a gain  of around \unit[9]{dB}. It points to the  Zenith and has a
half-power beamwidth (HPBW)  of $90^\circ$.  It is tuned  at a central
frequency of 3.8~GHz and a bandwidth of approximately \unit[500]{MHz}.
It  is associated with  a low  noise block  (LNB) which  amplifies the
signal  by approximately  \unit[60]{dB}  and lowers  down the  central
frequency  to  \unit[1.35]{GHz}.  The  antenna  and  the  LNB will  be
referred to LNBf  hereafter. A bias tee is inserted  after the LNBf to
both distribute the power supply to the LNB and transmit the RF signal
on  a \unit[75]{$\Omega$}  line.   The line  impedance  is adapted  to
\unit[50]{$\Omega$} by  a resistor bridge.  The  low-frequency part of
the spectrum  is filtered out  by a \unit[900]{MHz}  high-pass filter.
The adaptation electronics of \mbox{GIGAS61} is made partly with commercially
available device.  The power detector used, the Minicircuit ZX47-50 is
the encapsulated version of the  Analog Device AD8318. The rest of the
adaptation is carried out with  a custom made board.\\A first array of
seven detectors was installed at the Pierre Auger Observatory in April
2011.  The smooth operation and the results of this first test bed led
to an  extension by  54 more detectors  covering a  total instrumented
surface of  \unit[93]{$\rm km^2$}.   The first seven  LNBf are  of the
model GI301 made by Global Intersat  and the 54 units of the extension
are from WSInternational, model  DMX241.  The \mbox{GIGAS61} array is located
in the South-West part of the Pierre Auger Observatory.  Its footprint
is shown  in Fig.~\ref{fig:detector}-left.  Even if the  MBR signal is
expected unpolarized,  we fixed the polarization of  each antenna. Out
of  the 61 antennas,  33 have  a North-South  polarization, and  28 an
East-West one.\\ Several radio signals  in the C-band were detected in
coincidence  with Auger  EAS events  with  \mbox{GIGAS61}\cite{amber}.  These
detections validated the concept of the coincident radio detection and
were the  first detections  of EAS in  the C-band.  However,  such EAS
emissions in the microwave band  may have a different origin than MBR.
In  particular, the  signals were  detected  at distances  to the  air
shower axis of a few hundred meters only.  This feature is in favor of
the hypothesis of a beamed emission over an isotropic one as origin of
these signals.

\subsubsection*{GIGADuck}
\begin{figure}[!t]
  \centering
  \hspace*{-3ex}  
  \subfigure{\includegraphics[width=0.65\linewidth]{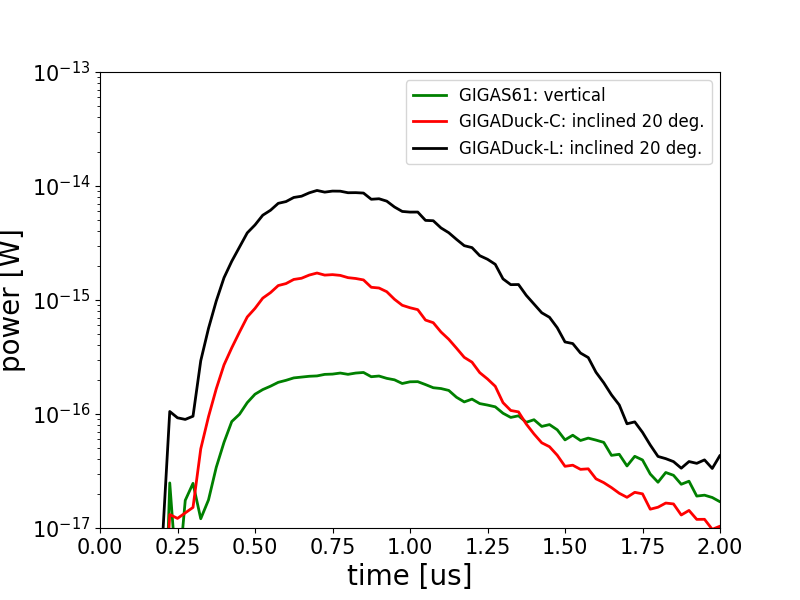}}
  \caption{Simulation  of the power  received from  a vertical  10 EeV
    shower  as a  function  of time  for  GIGAS and  \mbox{GIGADuck-C}
    detectors.}
  \label{fig:simexample}  
\end{figure}

The need to  improve our sensitivity at large  distance and to collect
more data led to the  design and installation of two optimized arrays,
in  the C-band  and  the L-band,  with  a higher  antenna  gain and  a
modified antenna orientation.  The array  is now composed of a central
detector pointing  to the Zenith with six  peripheral detectors tilted
by  20$\rm ^{\circ} $  in zenith  and with  azimuth adjusted  to point
towards  the   central  detector  (Fig.~\ref{fig:detector}-middle  and
right).   This configuration  increases the  overlap of  the detectors
field  of view  and enhance  the  probability to  obtain a  coincident
detection.   Indeed,   this  configuration  was   chosen  because  the
observation of a coincidence between two radio detectors would support
the  hypothesis of  an isotropic  emission.   \\As an  example of  the
improved performance  of GIGADuck, the simulation of  the radio signal
power produced by MBR emitted by  a vertical shower and detected by an
antenna  belonging  either  to  \mbox{GIGAS61}  or to  GIGADuck  at  a
distance of  \unit[750]{m} is shown  in Fig.~\ref{fig:simexample}.  In
the case of  this particular configuration of distance  and angle, the
signal collected by the tilted \mbox{GIGADuck-C} antenna is around ten
times larger, due  mainly to the higher gain and  the direction of the
main lobe.  In  the L-band, the signal is  increased by another factor
10,  due to  the quadratic  dependence of  the effective  area  of the
antenna  with  the   wavelength  (see  Eq.~\eqref{eq:aeff}).   Further
comparisons  are  shown   in  the  section~\ref{sec:simulation},  they
include calibrated value of  detector noise and realistic distribution
of the shower energies and arrival direction.

\paragraph{GIGADuck-C} 
In the C-band,  the antenna is a pyramidal  horn of \unit[15]{dB} gain
from the  A-Info company. It  increases the maximum  antenna effective
area by a  factor of six with respect to the  antennas of \mbox{GIGAS61}.  It
is protected  by a  thin radome  in plexiglass.  The  LNB is  a Norsat
8115F.   It was  chosen for  its low  noise figure  and has  a flatter
response in frequency with respect to \mbox{GIGAS61} horns, and thus a larger
bandwidth.

\paragraph{GIGADuck-L}
In  the L-band,  the sensor  is a  helicoidal antenna  with  a conical
copper grid  at its  base. It is  tuned to  be sensitive to  a central
frequency of  \unit[1.4]{GHz} and a gain of  \unit[15]{dB}. The sensor
is directly followed  by an electric surge protection  and a band-pass
filter (from \unit[1.1 to 1.4]{GHz})  to decrease the amplitude of the
signal  at  \unit[900]{MHz}  caused  by   the  GSM  band  and  the  SD
communication  system.  The  choice  of placing  a  filter before  the
amplifier is not optimal in terms  of noise figure but is necessary to
prevent the amplifier saturation.  The amplification board is composed
of two separated  amplifiers from the Avago company,  the MGA633P8 and
the MGA13116.  They are combined to  obtain a gain of around 50~dB. In
both GIGADuck  versions the adaptation  electronics was made  on custom
made board with discrete components. 

\section{Detector calibration}
\label{sec:calibration}	
EASIER  detectors   are  required  to  measure   faint  and  impulsive
signals. The widely used  figure of merit of the  sensitivity for such
detectors reads as
\begin{equation}
F  =  \frac{k_{\text{B}}  T_{\text{sys}} }{A_{\text{eff}} \sqrt{\Delta  \nu    \Delta t}},
\label{eq:sensitivity}  
\end{equation}
where  $F$ represents  the flux  resulting  from a  signal that  would
equate the noise fluctuations,  $T_{\text{sys}}$ stands for the system
noise equivalent  temperature (the sum of the  thermal noise collected
by the  antenna and  the electronics noise  added mainly by  the first
amplifier), $k_\text{B}$  is the Boltzmann  constant, $A_{\text{eff}}$
is  the  effective area  of  the antenna  (i.e.   the  portion of  the
incoming radio flux transformed into electrical power), and the square
root  term  is  the  amount   of  samples  over  which  the  noise  is
averaged. In simple cases, $\Delta \nu \Delta t$ is the product of the
bandwidth $\Delta \nu$ with a time  constant of a low pass filter, but
in cases  of transient signals,  the expected duration of  the signals
has to be used for $\Delta t$.  We detail first the calibration of the
sensor including  the measurement or  simulation of the  parameters in
Eq.~\eqref{eq:sensitivity}.   In  a   second  time  we  determine  the
calibration parameters of the adaptation stage of the signal chain.

\subsection{Sensor calibration}
\label{sec:calibrationsensor}
\subsubsection{Antenna effective area}
\begin{figure}[!ht]
 \centering
 \hspace*{-3ex}
 \subfigure{\includegraphics[width=0.65\linewidth]{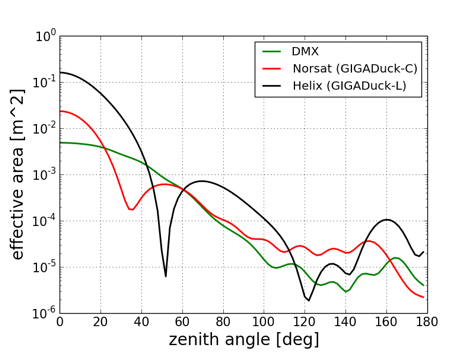}}
 \caption{Simulated     effective    area    for     \mbox{GIGAS61}    (DMX),
   \mbox{GIGADuck-C}(Norsat) and \mbox{GIGADuck-L} (Helix) antennas.}
 \label{fig:aeff}
\end{figure}

The effective  area for a  particular wavelength $\lambda$  is derived
from  the knowledge  of the  antenna gain  pattern  $G(\theta, \phi)$,
i.e. the gain of the antenna as a function of the direction:
\begin{eqnarray} \label{eq:aeff}
A_{\text{eff}} (\theta, \phi) = {{\lambda^2\ G(\theta, \phi)}\over{4 \pi} } 
\end{eqnarray}
The  gain pattern can  be either  measured or  simulated. It  has been
measured for the  antenna DMX241 from WS International  and for an ATM
horn  coupled to  a Norsat  LNB  in an  anechoic chamber  at the  IMEP
(Institut  de Microelectronique  Electromagnetisme  et Photonique)  at
Grenoble.   In  addition to  these  measurements,  the High  Frequency
Simulation Software (HFSS) from ANSYS~\cite{hfss} was used to simulate
the patterns of  the different antenna types, taking  into account the
setup of the sensors, such as  the presence of a radome. The simulated
effective area of the antenna used in the three setups are represented
in Fig.~\ref{fig:aeff}.
\subsubsection{System noise temperature}
\begin{figure}[!ht]
 \centering
 \hspace*{-3ex}
 \subfigure{\includegraphics[width=0.45\linewidth]{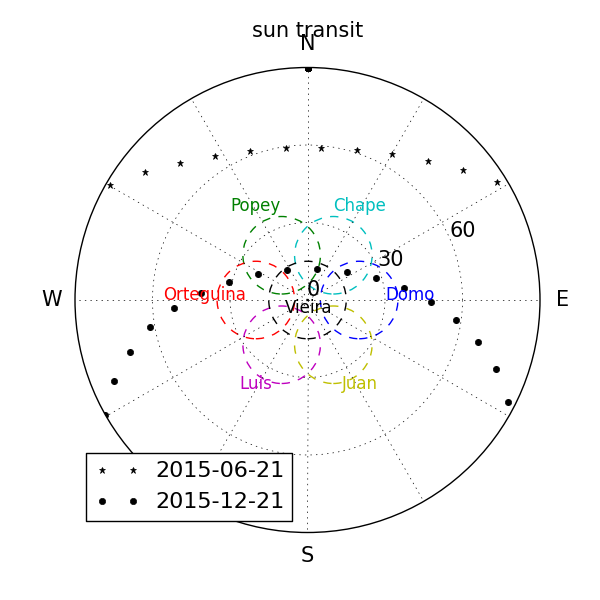}}
 \subfigure{\includegraphics[width=0.55\linewidth]{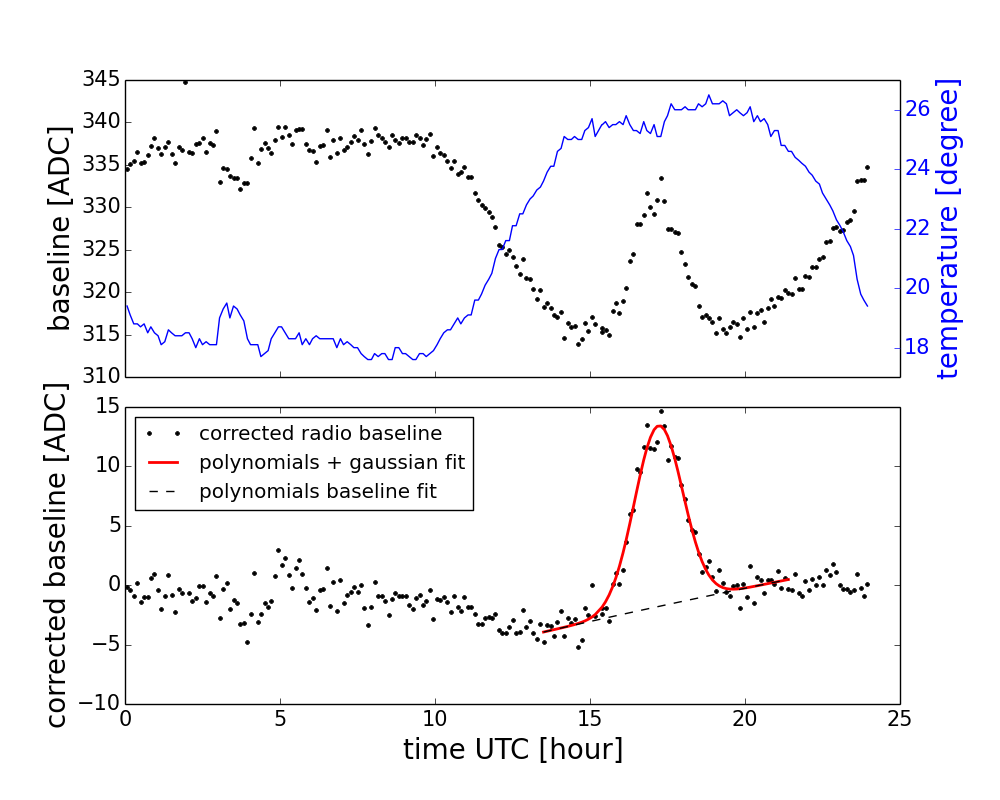}}
 \caption{Left: Sun transit for the two solstices in polar coordinates
   where the radial distance  represents the zenith angle. The azimuth
   is set to  0 for the East direction.  The colored circles represent
   the field of view of  the GIGADuck antennas.  Right: Example of the
   baseline during one  day. The top plot shows  the original baseline
   and the outside  temperature along the day.  The  bottom plot shows
   the baseline  once corrected from the  temperature dependence.  The
   fit result is also shown in red.}
 \label{fig:sunsim}
\end{figure}

\begin{figure}[!ht]
 \centering
 \hspace*{-3ex}
 \subfigure{\includegraphics[width=0.53\linewidth]{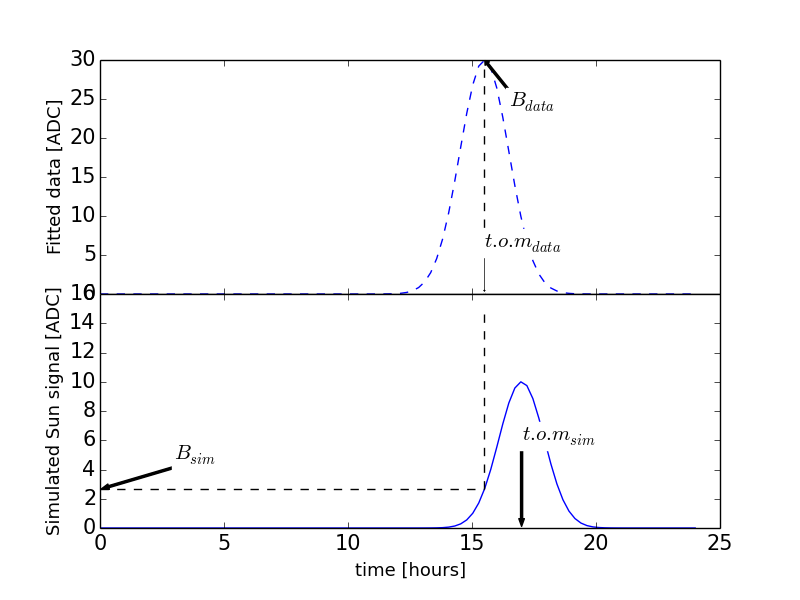}}
  \subfigure{\includegraphics[width=0.46\linewidth]{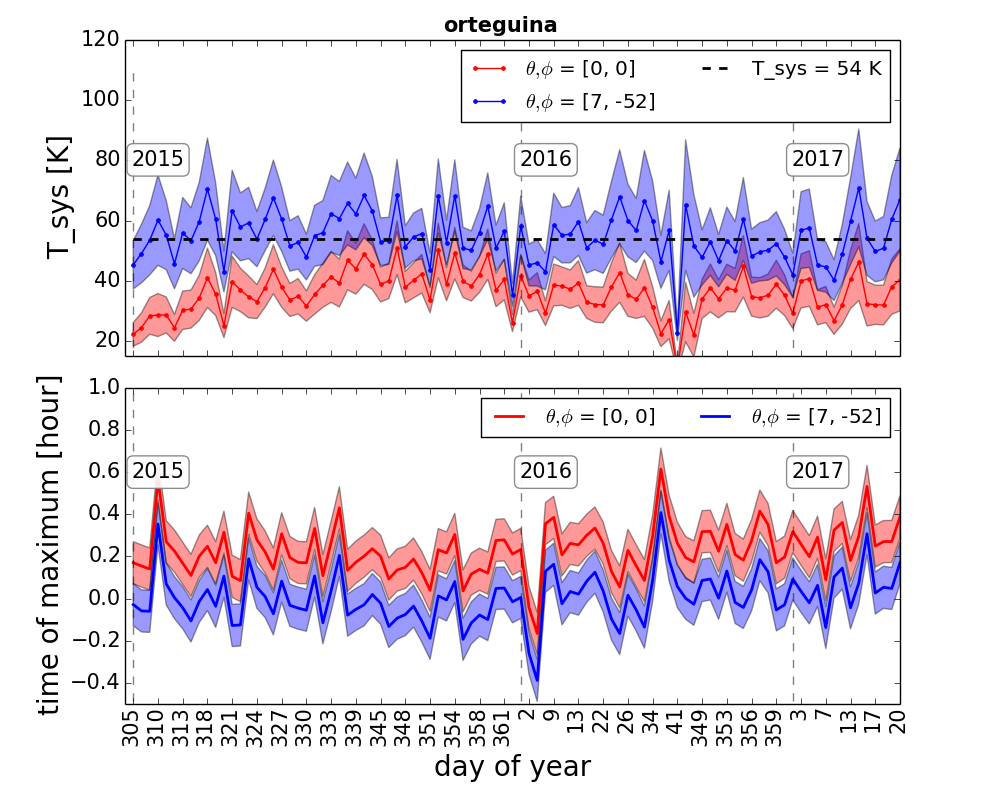}}
  \caption{Left:  Scheme with  the parameters  used in  the $\chi^{2}$
    function in \eqref{eq:chi2}.   Right: Temperature measured for one
    GIGADuck  detector (Orteguina)  with the  Sun signal  and  time of
    maximum.   We show together  the results  for the  nominal antenna
    orientation  (in   red)  and  the  one  retrieved   from  the  Sun
    observations (in blue).}
 \label{fig:GDtempres}
\end{figure}

The  system  noise  factor  is  defined  as  the  degradation  of  the
signal-to-noise ratio (SNR) along the signal chain stages  and can be
expressed with a system  noise temperature, $T_{\text{sys}}$. The main
contributions to the noise temperature are:
\begin{itemize}
\item  the antenna  temperature $T_{\text{ant}}  $: the  thermal noise
  emitted by  broad microwave  sources such as  the sky or  the ground
  collected by the antenna;
\item the electronics noise  $T_{\text{elec}}$: the noise added by the
  electronics  stage,   usually  dominated  by  the   first  stage  of
  amplification.
\end{itemize}
A radome  used to  protect an  antenna is a  source of  signal losses,
affecting  the  SNR  and  adding   up  a  contribution  in  the  noise
temperature.\\  To  estimate the  temperature  of  the three  detector
versions,  we have  applied  three different  methods.   A well  known
method to  measure a  temperature contribution from  a device  like an
amplification system  is the  $Y$ factor method.   It consists  in the
measurement of the  device output when it is  subject to two different
known  sources   of  noise.   In   the  case  of   \mbox{GIGAS61}  and
\mbox{GIGADuck-C} detectors,  the amplification system is  part of the
feed and cannot be isolated and tested separately. Hence, to apply the
$Y$  factor method and  produce a  stable noise  in the  detector, the
source has to  be a microwave emitting source that  covers most of the
antenna main lobe.  Having two references allows one to cancel out the
gain of  the system and  to extract the  noise.  Another method  is to
make use of  a natural microwave source like the  Sun as a calibration
source when it goes through the field of view of the antenna.  Lastly,
one can  also directly measure  the background radio power  and deduce
the noise temperature provided a precise knowledge of the system gain.
The $Y$  factor method was used  in a dedicated  measurement to obtain
\mbox{GIGAS61} detector  temperatures.  The second method  is used for
the  GIGADuck arrays  by  measuring  the Sun  flux  in the  monitoring
data. As it  will be described later, the Sun signal  was also used to
correct    for   the    pointing   direction    of   \mbox{GIGADuck-C}
antennas.  Finally the  direct method  is used  to measure  the L-band
setup system noise temperature.
\paragraph{GIGAS61 --}
For GIGAS61 detectors,  we apply the $Y$ factor  method to measure the
electronics  noise  temperature. The  measurement  took  place at  the
detector site.  The setup comprises the main components of the nominal
detectors,  namely  the  LNBf,  the  radome  and  the  power  detector
Minicircuit ZX47-50.   The antenna  was oriented consecutively  up and
down  and  the voltage  of  the power  detector  was  recorded with  a
portable  oscilloscope.   The   voltage  difference  between  the  two
measurements  is related  to a  difference of  power according  to the
calibration curve  of the power detector  (see Eq~(\eqref{eq:eqpd}) in
section~\ref{sec:calibrationadapt}).      The     electronics    noise
temperature $T_\text{elec}$ is computed with:
\begin{equation}
	T_{\text{elec}} = \frac{T_{\text{hot}} - YT_{\text{cold}}}{Y-1},
\end{equation}
where    $Y   =    P_{\text{hot}}/P_{\text{cold}}$,   $T_{\text{hot}}$
($T_{\text{cold}}$) is the antenna temperature when the antenna points
toward the  ground (the sky)  and $P_{\text{hot}}$ ($P_{\text{cold}}$)
are  the   corresponding  powers.  The  antenna   temperature  is  the
brightness  temperature of  the  surrounding sources  weighted by  the
antenna gain:
\begin{equation}
T_{\text{ant}} = \int_{\theta =  0}^{\theta = \pi}\int_{\phi = 0}^{\phi =2\pi} T_{\text{B}}(\theta, \phi) G(\theta,\phi) \sin{\theta} d\theta d\phi
\label{eq:tant}
\end{equation}
with   $T_{\text{B}}(\theta)$  the   brightness  temperature   in  the
direction  $\theta$.  We applied  the  formula~\eqref{eq:tant} with  a
brightness  temperature  profile  (found  in~\cite{otoshidanslacolle})
which ranges  from \unit[4]{K} in  the sky direction  to \unit[270]{K}
towards  the   ground.  Antenna  temperatures   of  $T_{\text{hot}}  =
\unit[260]{K}$  and $T_{\text{cold}} =  \unit[6]{K}$ are  obtained. It
yields to electronics  temperatures of $T^{\text{GSI}}_{\text{elec}} =
\unit[(114\pm    10)]{K}$    and    $T^{\text{DMX}}_{\text{elec}}    =
\unit[(97\pm  9)]{K}$.    Finally  we  add   the  antenna  temperature
$T_{\text{ant}} = T_{\text{cold}} =  \unit[6]{K}$ to obtain the system
noise temperature.
\paragraph{GIGADuck-C --}
Compared  to  GIGAS61  detectors,  GIGADuck  antennas  have  a  larger
effective  area which  make them  sensitive  to the  Sun flux.   Since
GIGADuck data are part of  the SD data stream including the monitoring
system, the radio baseline  is recorded every \unit[400]{s} with other
information  such as  the outside  temperature. We  use these  data to
search for the Sun signal and estimate the system temperature from it.
The position of the Sun in the sky is well known and the absolute flux
density in the frequency band is based on observations at the Nobeyama
Radio  Observatory  (NRO)  at \unit[3.75]{GHz}~\footnote{The  Nobeyama
  Radio  Polarimeters are  operated by  Nobeyama Radio  Observatory, a
  branch of  National Astronomical Observatory of  Japan}. Examples of
the  Sun  path  through  the   GIGADuck  C-band  array  are  shown  in
Figure~\ref{fig:sunsim}. Most  of the antennas (except  the antenna on
the stations called Juan and  Luis) have the Sun passing through their
field of view during the austral summer.  However, when the Sun is low
in  the sky  (during austral  winter time),  none of  the  antennas is
sensitive  to it.  Since all  GIGADuck antennas  point in  a different
direction, one  expects the  Sun to produce  a signal  with relatively
different  intensity and shifted  time of  maximum according  to their
orientation.  Indeed,  we use  both information to  constrain together
the  system  noise  temperature  and  the pointing  direction  of  the
GIGADuck antenna. \\
The increase of  power $P_\text{Sun}$ induced upon the  passage of the
Sun over the system noise power $P_\text{sys}$ in the antenna field of
view reads
\begin{equation}
  \label{eq:deltaP}
  \unit[\Delta  P]{[dBm]}  =  10\log_{10}\left(\frac{P_{\text{sys}}  +
    P_{\text{Sun}}(\theta_{\text{Sun}},\phi_{\text{Sun}})
  }{P_{\text{sys}}}     \right)    =    10\log_{10}\left(1+\frac{1}{2}
  \frac{F_{\text{Sun}}A_{\text{eff}}(\theta_{\text{Sun}},\phi_{\text{Sun}})}{k_\text{B}
    T_{\text{sys}}} \right),
\end{equation}
where  $F_{\text{Sun}}$  is  the  total  solar flux  measured  by  the
Nobeyama   Radio   Observatory~\cite{nobeyama}  at   \unit[3.75]{GHz},
$A_{\text{eff}}(\theta_{\text{Sun}},\phi_{\text{Sun}})    $   is   the
antenna effective area  for the given position of the  Sun in the sky,
and the factor $1/2$ is the polarization factor.
The radio baseline is  strongly correlated to the outside temperature,
but can  also be  affected by the  humidity in  a non trivial  way. We
first operate  a selection on the  dataset to isolate a  set of stable
days,  i.e.  the  days  when the  baseline  is not  affected by  other
parameters but the outside temperature.  \\Firstly, a time window of 8
hours around the time when the  Sun is expected to produce the highest
signal is removed temporarily.  Then  we reject the singular days when
the  baseline   RMS  is  lower   than  2~ADCu  (compared   to  20~ADCu
typically). These  low variations indicate  that the signal  chain was
faulty  at  that  time.   Days  with large  amplitude,  often  due  to
thunderstorm condition, are  removed by requiring baseline differences
over the day  lower than 200~ADCu. From this  data set, the dependence
of the  radio baseline with the  outside temperature is  fitted with a
linear function.   To improve  the selection of  stable days,  the day
with  the largest  residual is  removed and  the fitting  procedure is
repeated until  no residual larger  than 10~ADCu is found.   Then, the
time window which encompasses the  Sun contribution is restored in the
selected  days  and  the  complete  baselines are  corrected  for  the
temperature dependence.   The final step consists in  fitting the bump
induced by  the Sun flux  with a Gaussian  function and a  third order
polynomial.    An  example  of   the  radio   baseline  is   shown  in
Figure~\ref{fig:sunsim}-right before  the temperature correction (top)
and after (bottom).   \\The selection and fit procedure  are tested by
introducing fake  signals to  mimic the Sun  contribution in  the real
baselines of the  antennas oriented towards the South  (Luis and Juan)
thus  insensitive to  the  Sun.   Signals with  a  Gaussian shape  are
introduced  with  various amplitude  and  time  and are  reconstructed
according to the  method described above.  The uncertainty  due to the
limited knowledge  of the  baseline amounts to  $\pm$\unit[4]{ADCu} on
the amplitude of  the peak and to $\pm$\unit[12]{minutes}  on the time
of  maximum.  The  spread of  the result  of the  fit is  found  to be
$\pm$\unit[5]{ADCu} and $\pm$\unit[6]{minutes}.  \\The goal is to find
the best parameters  to describe the system noise  temperature and the
pointing direction  given the observed amplitude and  time of maximum.
We simulate the signal induced by  the Sun microwave flux for a system
temperature  from \unit[30 to  120]{K} with  \unit[1]{K} step  and for
angles  $\Delta \theta  \in$ [0$^{\circ}$;  20$^{\circ}$]  and $\Delta
\phi \in  $[0$^{\circ}$; 180$^{\circ}$]  around the nominal  angle for
the set  of days selected  in the aforementioned procedure.   For each
set  of  input parameters  ($T_\text{sys}$,  $\Delta \theta$,  $\Delta
\phi$), the baselines  in ADCu are computed.  The  best parameters are
found by minimizing the following $\chi^2$:
\begin{equation}
	\chi^2   (T_\text{sys})  \rvert_{\Delta   \theta,\Delta  \phi}
        =\sum_{i}         \frac{         (t.o.m._\text{i,sim}        -
          t.o.m._\text{i,data})^2}{\sigma_\text{t.o.m}^2}             +
        \frac{(B_\text{i,sim}           -          B_\text{i,data})^2}
        {\sigma_{\text{B}^2}}
	\label{eq:chi2}
\end{equation}
where each day is labeled with  the index $i$, $t.o.m.$ stands for the
time of the maximum in data and simulation, $B_\text{{i,data}}$ is the
maximum   of   the  fitted   signal   in   ADCu   in  the   data   and
$B_\text{{i,sim}}$ the signal  in the simulation taken at  the time of
the   maximum   measured   in    the   data   (see   the   scheme   in
Figure~\ref{fig:GDtempres}-left).   The   result  are  given   in  the
table~\ref{tab:temptab}.  Angular deviations from the nominal position
are found to be at  most \unit[14]{$^\circ$} (in angular distance) and
temperatures range  from \unit[54]{K}  to \unit[61]{K}. An  example of
the temperature measured for each day  in the data set and the time of
maximum    compared   to    the    simulated   one    is   shown    in
Figure~\ref{fig:GDtempres}-right.
\begin{center}
\begin{table}[t]
\caption{Results   of  fits,   superscript  and   subscript   are  the
  statistical and systematics uncertainty respectively.}
\vspace{3mm}
\centering
\resizebox{\linewidth}{!}{
\begin{tabular}{cccc}
\toprule station name & Popey  & Orteguina & Domo \\ \midrule original
orientation $\theta  / \phi$ &  20 / 120 &  20 / 180  & 20 / 0  \\ new
orientation       $\theta^{\text{stat}}_{\text{sys}},      \phi^{\text{stat}}_{\text{sys}}$      &
25$^{+1/-1}_{+1/-1}$  /  116$^{+1/-1}_{+1/-2}$  &23$^{+1/-1}_{+3/-2}$,
170$^{+1/-1}_{+1/-1}$              &             33$^{+1/-1}_{+1/-1}$,
12$^{+5/-5}_{+6/-5}$\\ System Temperature $T^{\text{stat}}_{\text{sys}} $ & 61$^{+2/
  -2}_{+12/-10}$  &   54$^{+2/-1}_{+12/-7}$  &  58$^{+2/-3}_{+8  /-9}$
\\ \bottomrule
\end{tabular}
}
\label{tab:temptab}
\end{table}
\end{center}

\paragraph{GIGADuck-L --}
The L-band sensors are also sensitive to the Solar flux.  Thirty daily
baselines are  overlaid in Fig.~\ref{fig:GDLbaseline_BW}  and exhibits
the Sun  passage (around \unit[18]{h}) but also  other modulation (for
instance around \unit[0]{h}  or \unit[7]{h}). These modulations, whose
origin  may  be the  positioning  satellite  signal,  prevent us  from
quantifying the Sun contributions in the same way as above. Hence, the
noise  temperature  is deduced  from  the  direct  measurement of  the
baseline, simply by  dividing the measured power by  the total gain of
the system. This method requires a precise calibration of the absolute
gain   of   the   detector   which   was  performed   prior   to   the
installation. The  amplifier is pre-terminated and its  gain and noise
temperature  could  be measured  respectively  with  a Vector  Network
Analyser and a Noise figure  meter.  In the C-band this measurement is
made difficult by the use  of LNBf and the impossibility to disconnect
the  amplification stage  from the  feed waveguide.  The  system noise
temperature is measured for all seven L-band detectors, it ranges from
\unit[94]{K} to \unit[145]{K}.

\subsubsection{Sensor bandwidth}
\begin{figure}[!ht]
 \centering
 \hspace*{-3ex}
 \subfigure{\includegraphics[width=0.49\linewidth]{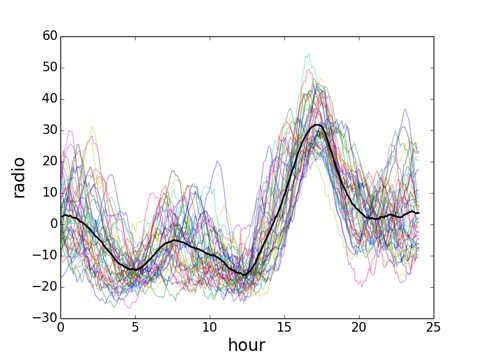}}
	 \includegraphics[width=0.49\linewidth]{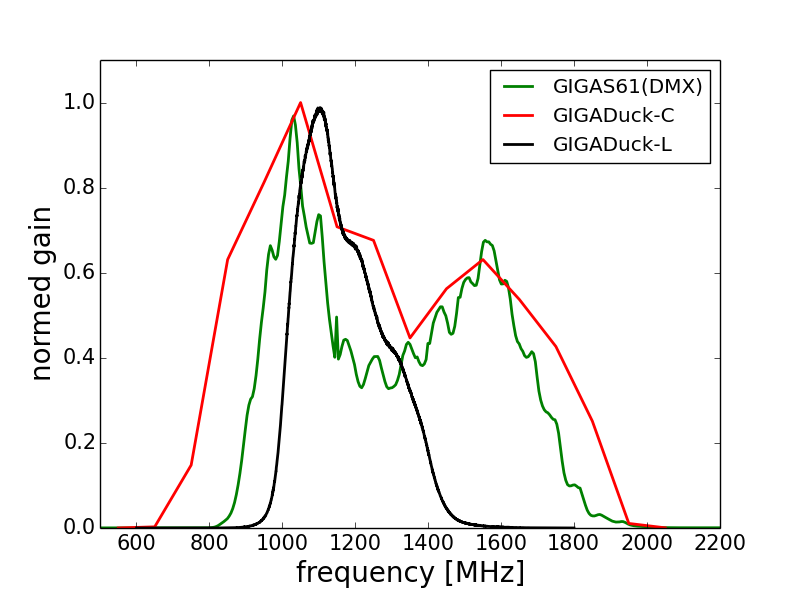}
 \caption{Left:    30    daily    baselines   for    Jorge    detector
   (GIGADuck-L). Right:  Normalized gain  of the three  mentioned LNBf
   after the frequency downconversion.  The thick blue and green lines
   are the  average over several  detectors. Only one  measurement was
   performed with the Norsat LNBf (in red). }
 \label{fig:GDLbaseline_BW}
\end{figure}

The absolute  gain of  the RF part  which includes the  amplifier, the
bias tee,  the cables  etc., does not  enter directly in  the detector
sensitivity,    but     the    frequency    bandwidth     does    (see
Eq.~\eqref{eq:sensitivity}).  The normalized gains of the LNB used for
\mbox{GIGAS61} (DMX241) and \mbox{GIGADuck-C}  and the \mbox{GIGADuck-L} are represented in
Fig.~\ref{fig:GDLbaseline_BW} and the  effective bandwidth is computed
according to:
\begin{equation}
  \Delta \nu = \frac{1}{G_{\text{max}}} \int G(f) df
\end{equation}
The obtained effective bandwidths  for \mbox{GIGAS61} detectors are \unit[437
  $\rm \pm$  30]{MHz} and \unit[445  $\rm \pm$ 56]{MHz} for  the GI301
and the  DMX241 respectively.  As for the  \mbox{GIGADuck-C}, a  bandwidth of
\unit[750]{MHz} is measured for the Norsat LNB, and finally an average
of \unit[250]{MHz} is found for the \mbox{GIGADuck-L} LNAs.

\subsection{Electronics calibration}
\label{sec:elec}
We describe  here the functioning of the  adaptation electronics.  The
first part of this section is  dedicated to the study of steady signal
of the  adaptation needed  to describe the  baseline level,  while the
second  part describes the  time response,  necessary to  simulate the
full signal  chain. 
\subsubsection{Response to steady signals}

\label{sec:calibrationadapt}
The adaptation electronics  is composed of the power  detector and the
adaptation  board. The power  detector output  voltage $V_{\text{pd}}$
was calibrated  in laboratory using  a noise waveform.  The  noise was
produced  using the  output of  an actual  LNBf placed  in front  of a
microwave absorber  to obtain the same  spectrum as in  the data.  The
input  power   $P_{\text{in}}$  was  varied   with  attenuators.   The
power-voltage characteristic reads:
\begin{equation}
  V_{\text{pd}}  [\text{V}]  =  -0.0234 P_{\text{in}}  [\text{dBm}]  +
  \text{offset}_1,
\label{eq:eqpd}
\end{equation}
$\text{offset}_1 $ is  the voltage offset of the  power detector.  The
power detector voltage  is then amplified by a factor  4.2 to obtain a
final power  dynamics of \unit[20]{dB}  over the \unit[2]{V}  swing of
the SD  acquisition.  An offset was  designed to be  adjustable on the
adaptation  board to  make  up  for the  differences  of the  detector
gains. The overall conversion from the input power to the ADCu is:
\begin{equation}
  \text{P[ADCu]} = 50.2 P_{\text{in}} [\text{dBm}] + \text{offset}_2,
\label{eq:eqcalibration}
\end{equation}
where $\text{offset}_2$ accounts for the total offset.

\subsubsection{Response to impulsive signal}
\label{sec:elecimpulsive}
\begin{figure}[!ht]
 \centering
 \hspace*{-3ex}
 \subfigure{\includegraphics[width=0.7\linewidth]{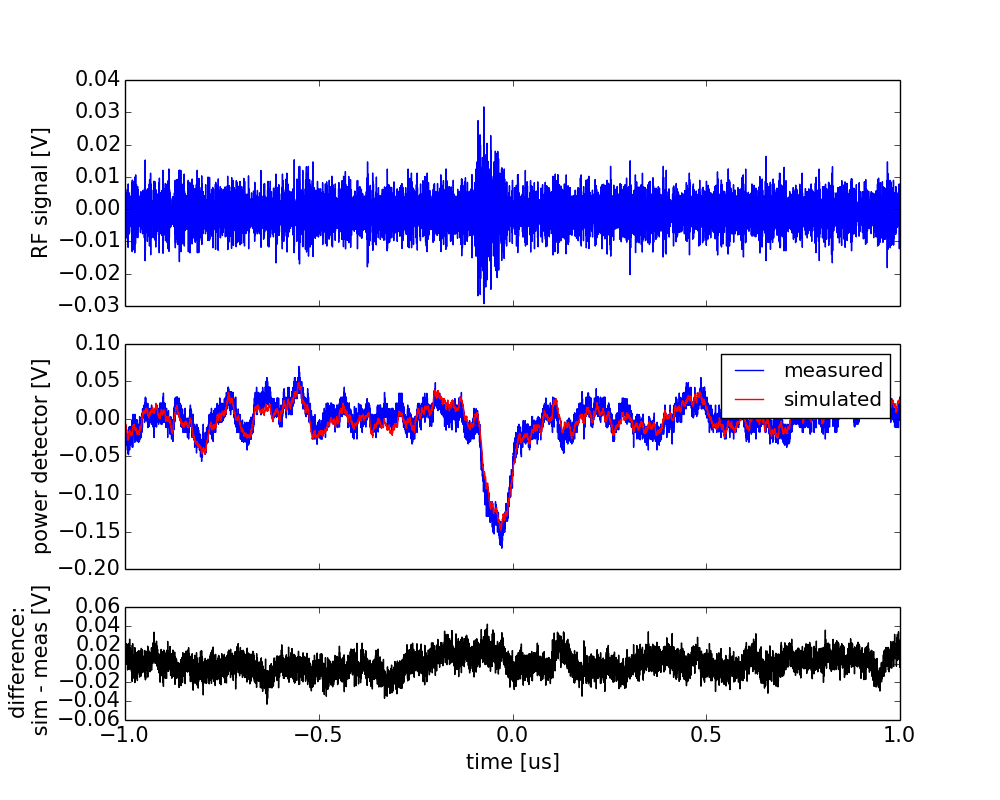}}
 \caption{Example of RF and  power detector waveforms.  The top pannel
   shows the  RF waveform, the  middle pannel the waveforms  after the
   power  detector: the  blue  one is  a  measurement and  the red  is
   simulated from  the waveform  on the top  pannel.  The  lower panel
   show the difference of the power detector waveforms.}
 \label{fig:powerdetsim}
\end{figure}

\paragraph{Power detector --}
To understand the power detector response to impulsive signals, we set
a detection chain  in the laboratory composed of a  LNBf followed by a
power  detector.  An  impulsive  and  high frequency  (HF)  signal  is
produced  by  the spark  of  an  electronic  lighter.  The  signal  is
recorded simultaneously after the LNBf and after the power detector by
a  fast  oscilloscope.   An  example  of these  signals  is  shown  in
Fig.~\ref{fig:powerdetsim}.   We  can  therefore  build  a  method  to
reproduce the  power detector output from  a HF signal.   We find that
the power  detector output  is well reproduced  when one  performs the
convolution  of  the  HF  signal  in dBm  (logarithmic  unit)  and  an
exponential function with a decay constant $\tau$:
\begin{equation}
  V_{\text{PD}}(t) = k_{1} \int_{t>0}P_{\text{dBm}}(u)\exp{\left(\frac{t-u}{\tau}\right)}du + k_{2}
  \label{eq:convolution}
\end{equation}
The   factor  $k_{1}$   is   fixed  to   the   conversion  factor   in
Eq.~\eqref{eq:eqpd},  $  k_{2}$  is  a  floating offset  and  $\tau  =
\unit[6.3]{ns}$ was found to provide the best fit to the data.
\footnote{The  first seven  detectors of  \mbox{GIGAS61} have  a  longer time
  response  $\tau_{\text{capa}} =  \unit[41.5]{ns}$ due  to  an output
  capacitor  present by  default  in the  power  detector ZX47-50  and
  removed in the following version of EASIER.}

\begin{figure}[!ht]
 \centering
 \hspace*{-3ex}
 \subfigure{\includegraphics[width=0.65\linewidth]{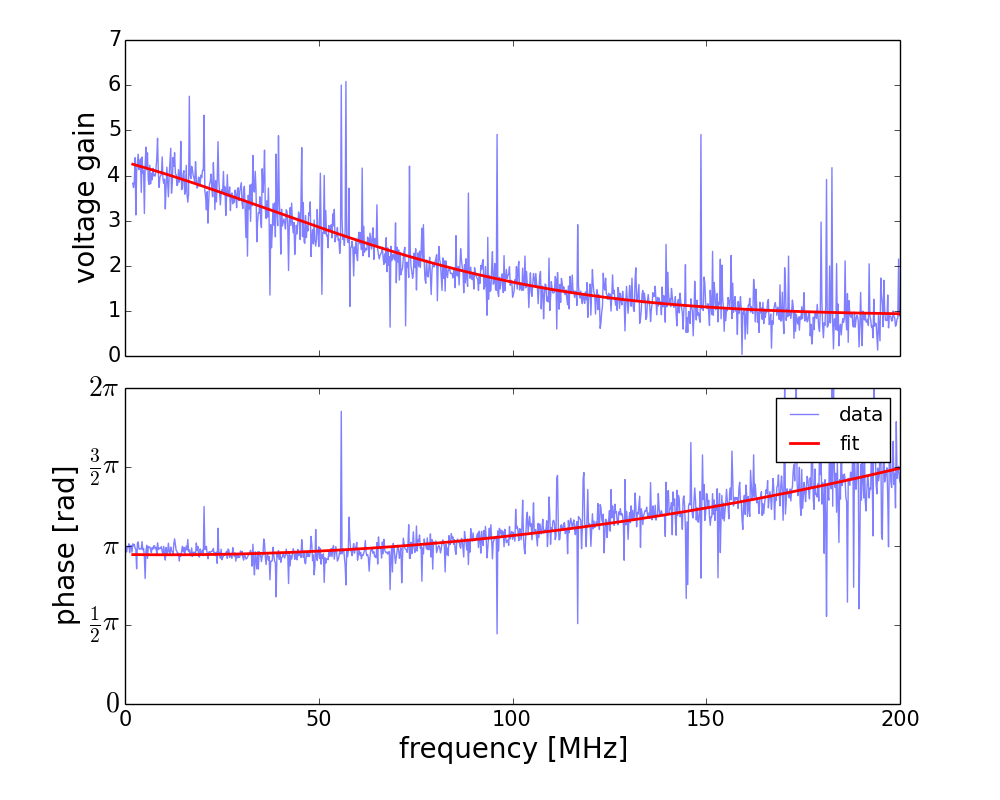}}
 \caption{Measurement and fit of the gain and phase of the adaptation board.}
 \label{fig:board}
\end{figure}

\paragraph{Adaptation board --}
To measure  the response  of the  adaptation board, we  add it  to the
calibration setup  described in  the previous paragraph.   We recorded
simultaneously  the input of  the board  and its  output. We  find the
board response  by measuring the transfer function $\tilde{H}(f)$  in the frequency
domain:
\begin{equation}
  \tilde{H}(f)                                                        =
  \frac{\tilde{V}_{\text{out}}(f)}{\tilde{V}_{\text{in}}(f)}.
\end{equation}
The   gain  and   the  phase   of   the  board   are  represented   in
Fig.~\ref{fig:board}.  The  time   response  is  obtained  by  Fourier
transformation. \\The last part of  the chain, the Auger SD front end, is
simulated   with  a   low-pass   filter  with   $   f_{\text{cut}}  =
\unit[20]{MHz}$ and by sampling in time and amplitude.

\section{Expected performances}
\label{sec:simulation}
This  section   is  dedicated   to  presenting  the   performances  of
\mbox{GIGAS61} and  GIGADuck detectors in terms of  EAS detection.  In
the section~\ref{sec:mbrsim}  we detail  the method we  implemented to
estimate the  flux from an EAS.   In the section~\ref{sec:eventnumber}
we apply this method to estimate  the expected number of events in one
year of operation of \mbox{GIGAS61} and GIGADuck detectors.
\subsection{Simulation of the MBR signal from EAS}
\label{sec:mbrsim} 
The longitudinal profile of EAS is parameterised with a Gaisser-Hillas
function~\cite{gh} characterising the number of charged particles at a
certain depth.  The mean values and  RMS of the parameters used in the
Gaisser-Hillas  function  are  first  tabulated for  energies  between
\unit[10$^{17.5}$  and   10$^{21}$]{eV}.   A  randomisation   is  then
performed when generating  an event using a Gaussian  function for all
the parameters except for the depth of first interaction which follows
an  exponential  distribution.  Starting  from  the first  interaction
point,  high in  the atmosphere,  the number  of primary  electrons is
calculated in grammage steps  of \unit[$\Delta X$ = 2.5]{g cm$^{-2}$}.
At  each step,  the mean  energy  deposit per  particle is  calculated
following a parameterisation at \unit[1]{MeV} given in~\cite{nerling}.
The  lateral  distribution function  of  the  electrons  in the  plane
orthogonal to the shower axis  is taken as an NKG function~\cite{nkg1,
  nkg2}.\\The estimation  of the  flux of MBR  photons emitted  by the
ionisation electrons and received at ground is based on the derivation
presented  in~\cite{imen2016}.  It accounts  for the  MBR differential
cross section obtained in~\cite{MBRXsec} and the time evolution of the
shower plasma  as the ionisation  electrons get attached or  see their
energy shifted as they undergo ionisation or excitation reaction.  The
flux folded  with the antenna  effective area and integrated  over the
frequency  bandwidth yields  the power  envelope  of the  signal as  a
function of the time at the receiver (an example of the power envelope
is shown  in Figure~\ref{fig:yieldnormal}-left).  Following  the model
in~\cite{imen2016}  we  find a  spectral  intensity of  \unit[$\mathrm
  2\times10^{-26}]{W   \  m^{-2}   \   Hz^{-1}}$  for   a  shower   of
\unit[$10^{17.5}$]{eV}  observed  at  \unit[10]{km}.  This  estimation
allows for  the comparison with  other MBR studies. For  instance, the
same reference shower in the  same conditions would produce a spectral
intensity of \unit[$\mathrm 2.77\times10^{-24}]{W \ m^{-2} \ Hz^{-1}}$
according to the results of SLAC T471~\cite{Gorham}, the original beam
test. We introduce  a scale factor $R$ based on  the comparison of the
reference shower,  with $R=1$ for the  model that we  used and $R=140$
for the  SLAC T471 assumption. The  parameter $R$ is used  in the next
section to  assess the performance  of the detectors.\\To  account for
the detector response, the voltage  deduced from the power envelope is
multiplied  with a noise  waveform produced  according to  the spectra
measured  and presented  in  section~\ref{sec:calibrationsensor}.  The
resulting waveform is the simulation  of the RF voltage induced at the
output of the  antenna by the EAS.  A noise  waveform is produced with
the same spectrum, but the average power is normalized with the system
noise temperature.  We add the  two waveforms to emulates the total RF
voltage.  The adaptation electronics is then simulated as described in
section~\ref{sec:elecimpulsive} to obtain a waveform in ADCu.
\subsection {Expected event rate}
\label{sec:eventnumber}
\begin{figure}[t!]
\centering \includegraphics[width=0.49\textwidth]{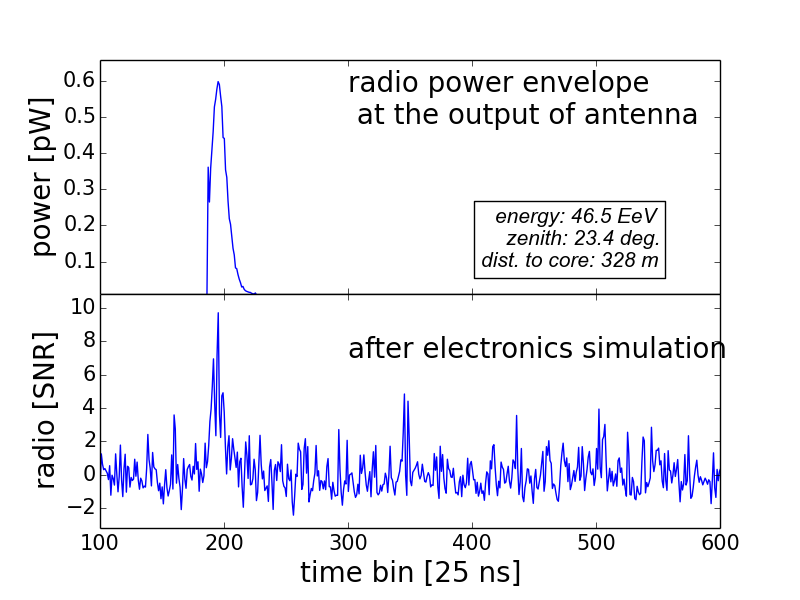}
\includegraphics[width=0.49\textwidth]{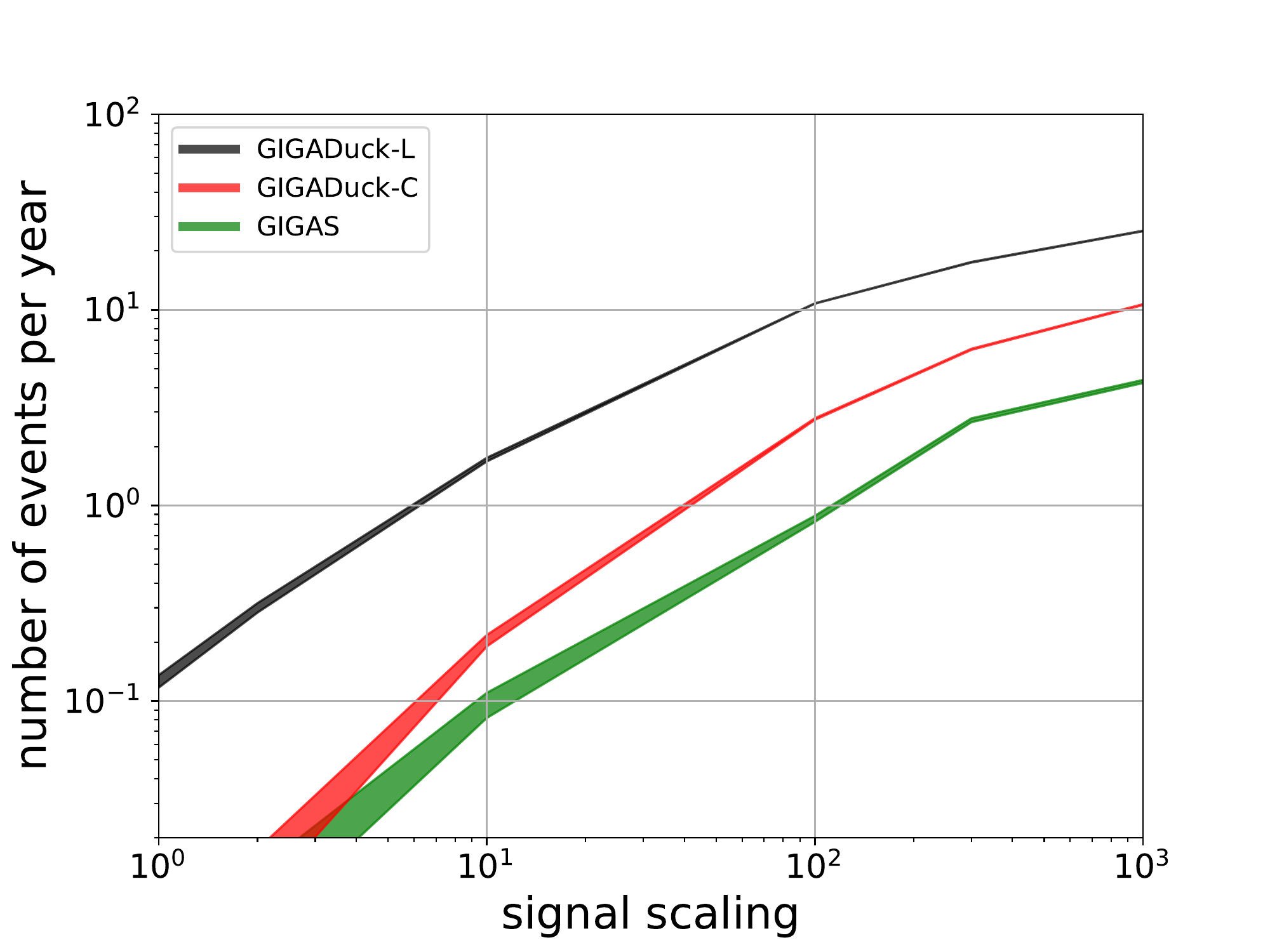}
\caption{Left: Example  of a simulated  radio signal at the  output of
  the antenna  for a scale  factor of 10,  and when combined  with the
  full electronics  simulation.  Right:  expectation of the  number of
  events  per  hexagon  per  year  as  a function  of  the  MBR  scale
  factor.   The  factor  for   SLAC  T471   is  $R$=140   (with  $R$=1
  corresponding to~\cite{imen2016} ).}
\label{fig:yieldnormal}
\end{figure}

For  a scale  factor $R$,  the number  of expected  events for  a time
period $\Delta T$ and for an  area labeled $S$ inside which the shower
core position is $x,y$ reads as:
\begin{equation}
	\mu(R) = J_0\int_{>E_0}dE~f(E) \int_{\Delta \Omega}d\Omega~\cos{\theta}\int_{S}dxdy~\int_{\Delta T}dt~\epsilon(E,\theta,\phi,x,y;R),
\end{equation}
where  $\epsilon$, the  detection  efficiency, is  estimated with  the
simulations  described  below.   The  energy  $E$  of  the  shower  is
generated randomly  following the energy  spectrum in the  range above
the so-called ankle energy which  can be parameterised between E$_0$ =
4$\times$10$^{18}$~eV      and     3$\times$10$^{20}$~eV     according
to~\cite{schulz}:
\begin{equation}
J(E;   E  >   E_0)   = J_0 f(E) = J_0   E^{-\gamma_2}   \left(1+  \exp   \left({
  \frac{\log_{10}E          -          \log_{10}E_{1/2}}{\log_{10}W_c}
}\right)\right)^{-1},
\end{equation}
where  $J_0$ is  a flux  normalisation factor  and the  spectral index
above the ankle $\gamma_2$ is 2.63. The term $\log_{10}E_{1/2}$ is the
energy at which the flux has  dropped to half of its peak value before
suppression, and $\log_{10}W_c$ is  its associated steepness. They are
fixed to  19.63 and  0.15 respectively. \\  Shower cores  are randomly
generated over a surface covering  an Auger hexagon, while the arrival
directions  $\theta$ and  $\phi$ are  randomly generated  to guarantee
uniformity  in terms  of  $\phi$ and  $\sin^2{\theta}$ (with  $\theta$
limited  to  60$^\circ$).  For  the  three detectors,  \mbox{GIGAS61},
\mbox{GIGADuck-C} and \mbox{GIGADuck-L}, we  simulate 5000 proton showers.  For each
shower we compute the MBR power  at the seven antennas of the hexagon.
Scale factors  $R$ from 1 to  1000 are applied and  the electronics is
then  simulated  ten  times  for  each $R$.   The  radio  waveform  is
transformed  in SNR  unit according  to: $P  [\text{SNR}] =  \frac{P -
  <P>}{\text{RMS}(P)}$   (see   Fig.~\ref{fig:yieldnormal}-left).   We
apply simple  selection criteria on  these processed data.   We select
events with a  waveform that passes a  threshold of SNR = 5  in a time
window of  \unit[1]{$\mu$s} around the expected time  of maximum.  The
expected number  of events for one  equipped hexagon within  a year of
data taking is  shown in Figure~\ref{fig:yieldnormal}-right, where the
abscissa  axis  is  the  scale factor.   The  initial  implementation,
\mbox{GIGAS61},  is already  sensitive to  the level  of  intensity as
measured by  SLAC T471 ($R = 140$)  but would observe only  one or two
events.  For the same scale  factor, this number increases by a factor
of  3  with  \mbox{GIGADuck-C}.   The  best performances  are  obtained  with
\mbox{GIGADuck-L} with possible detection down to scale factor of around $R =
10$.\\Several improvements in the  analysis would help to identify MBR
and are worth noting here.   The basic event selection can be improved
using digital filtering.  Such  analysis will particularly enhance the
long  duration signal  (a  few  $\mu$s) as  expected  for MBR  signal.
Furthermore,  in  contrast  to  the  geosynchrotron  emission  or  the
Askaryan effect, also present at  GHz frequencies, the MBR emission is
isotropic,  this gives the  possibility to  identify it  by requesting
that the  GHz radiation is  detected at large distances.   A plausible
selection criteria would  thus be performed on the  number of stations
that detected a  radio signal in coincidence with  the same shower. By
requesting  at least  two  stations  spaced on  the  regular SD  array
(1500~m   spacing),   one  would   discard   emissions  arising   from
geosynchrotron  or Askaryan  effects  (the expected  signals of  which
expand  over  a  few   hundred  meters  only).   Indeed,  the  antenna
orientation of GIGADuck antennas  was chosen optimized the coincidence
probability.  \\However, even with these methods, if the MBR intensity
is at the level  of the reference model~\cite{imen2016}, its detection
with the presented instruments is  hardly possible with only 0.1 event
expected per hexagon per year  at best.  To improve further the signal
to  noise  ratio,  other  experimental  techniques,  like  cryo-cooled
detectors, should be  considered. Note that the estimation  of the MBR
flux  is  delicate and  if  numerous  processes  are already  included
in~\cite{imen2016}  the  conclusions  on theoretical  predictions  are
uncertain  justifying the experimental  prospection being  carried out
with \mbox{GIGAS61} and the GIGADuck detectors.

\section{Conclusion}
GIGAS61 and its successors  \mbox{GIGADuck-C} and \mbox{GIGADuck-L} are designed for
the detection of MBR produced  by EAS in the microwave frequencies and
integrated in a Pierre Auger Observatory surface detector. This design
was proven to be efficient since GHz signals associated to air showers
were observed  with the first implemented array  \mbox{GIGAS61}.  However the
origin of such signal could not be attributed unambiguously to the MBR
and detectors with an enhanced sensitivity (\mbox{GIGADuck-C} and \mbox{GIGADuck-L})
were installed.   \\ We  have demonstrated the  good operation  of the
installed detectors  and we performed a calibration  of the parameters
useful to describe their sensitivities, namely the effective area, the
noise  temperature   and  the   bandwidth.   The  simulation   of  the
electronics,  and especially  the response  to short  pulses  was also
studied and shown  to be well understood.  \\The  performance of these
detectors  was   examined  under  various  assumptions   for  the  MBR
intensity, and with  a simulation of the detector  chain.  We verified
the improvement  of performance obtained with  the GIGADuck detectors.
For  the  most  sensitive  array (\mbox{GIGADuck-L}),  the  number  of
expected events is of the order  of 15 per year for one equipped Auger
hexagon when the original estimation from SLAC T471 is assumed.  While
the expectation from the most recent model of MBR emission is still be
out  of reach, \mbox{GIGAS61}  and GIGADuck  detectors is  able to  probe $in
\ situ$  the flux  of MBR over  two frequency bands.   \\The detectors
have been operating and accumulating data since 2011 for the first one
and since end of 2016 for the last installed one.  The analysis of the
data to search for MBR  signals or derive upperlimits on the intensity
are being now carried out and will be presented in a future paper.

\section*{Acknowledgement}
We gratefully acknowledge the very  fruitful exchanges we had with all
of our colleagues in the Auger  collaboration and the use of all Auger
facilities from hardware  to software including access to  a subset of
shower  data. We are  also deeply  indebted to  the commitment  of the
observatory staff  whose strong support in  constructing and deploying
the  LSD  prototype was  extremely  appreciated  \\We acknowledge  the
support of  the French  Agence Nationale de  la Recherche  (ANR) under
reference ANR-12- BS05-0005-01.

\bibliographystyle{elsarticle-num}
\bibliography{easier}

\end{document}